\documentclass[a4paper, 12pt]{article}
\usepackage{graphicx}
\usepackage{latexsym}
\usepackage{amsmath}
\usepackage{amsfonts} \usepackage{amssymb}
\usepackage[latin1]{inputenc}
\usepackage{amsmath}
\usepackage{epsfig}

\newcommand{\be}{\begin{equation}}
\newcommand{\ee}{\end{equation}}
\newcommand{\bea}{\begin{eqnarray}}
\newcommand{\eea}{\end{eqnarray}}

\begin{titlepage}
\author{Fernando de Felice \footnote{defelice@pd.infn.it}\\
{}
\\
{Dipartimento di Fisica  Universit\`{a} di Padova, Italy}\\
{I.N.F.N. Sezione di Padova,}\\
}
\title{Naked Singularities, Cosmic Time Machines and Impulsive Events}

 \end{titlepage}
\begin{document}
\maketitle

\begin{abstract}
{\it Continued gravitational collapse gives rise to curvature 
singularities. If a curvature singularity is globally naked then the 
space-time may be causally future ill-behaved admitting closed time-like 
or null curves which extend to asymptotic distances and generate a Cosmic
Time Machine (de Felice (1995) Lecture Notes in Physics {\bf 455}, 99).
The conjecture that Cosmic Time Machines give rise to high energy impulsive  
events is here considered in more details.}
\end{abstract}


\setlength{\baselineskip}{21pt}
\section{Introduction}

The outcome of naked singularities as result of
 gravitational collapse is still matter of debate. They have far reaching consequences; their 
space-time can be causally ill-behaved and
they may be sources of cosmic events with anomalous high energy. 
If they existed it would be legitimate to invoke the validity of a 
theorem due to Clarke and de Felice (1984)  which states 
that a generic strong-curvature naked singularity would give rise to a  
Cosmic Time Machine (CTM). 
A Cosmic Time Machine is a  space-time which is 
asymptotically flat and admits closed non-spacelike curves which extend to future infinity. 
Here I shall first recall the properties of a naked singularity and in particular those of a spinning
one then will illustrate what a Cosmic Time Machine is. 
Aim of this paper is to better motivate an earler 
conjecture (de Felice, 2004) according to which a CTM may be source of
fast varying and highly energetic events like Gamma Ray Bursts (GRB).  

\section{Naked Singularities} 

A naked singularity is the outcome of a continued gravitational collapse
when no event horizon forms hiding the singularity to the asymptotic region. 

A distinctive feature of a generic singularity 
is that of being infinitely red-shifted 
with respect to any of the non singular space-time points except possibly a 
set of measure zero. 
Since no physical influence reaches infinity from the singularity, then it is justified to assume
the existence of a regular flat (past and future) infinity after the formation of the singularity.

A further and indeed  most important feature of a naked singularity is that of 
approaching a black-hole state. There are various indications  as shown for example 
in (de Felice, 1975; 1978) that a naked singularity, specifically a spinning one, tends to 
become a black hole as 
result of its interaction with the surrounding medium. This implies that  
observable processes which take place nearby a naked singularity fade away, 
because of a growing red-shift,
in a finite interval of the observer's  proper time. 
This property is crucial to sustain the conjecture about the nature of strong 
impulsive cosmic events as I will illustrate next.
If a naked singularity decays into a black hole, then the latter  
will likely be of a Kerr type. Moreover when a naked singularity is close to become a 
Kerr black hole then it becomes of a Kerr type itself.

The properties of a Kerr naked singularity have been extensively investigated in the late seventies
(Calvani and de Felice, 1978; de Felice and Calvani, 1979) hence I will recall them briefly.
In Boyer and Lindquist coordinates, Kerr metric reads:
\begin{eqnarray}
\label{metric}
ds^2&=&-\left(1-\frac{2Mr}\Sigma\right)dt^2-\frac{4Mar\sin^2\theta}\Sigma dt d\phi+
\frac A\Sigma\sin^2\theta d\phi^2\nonumber\\
&+&\frac\Sigma\Delta dr^2+\Sigma d\theta^2
\end{eqnarray}
where $M$ is the mass of the metric source, $a$ its specific angular momentum\footnote{
 I use geometrized units, i.e. $c=1=G$, $c$ and $G$ being respectively the velocity of light 
in the vacuum 
and the gravitational contant.} and the functions $\Delta$ , $\Sigma$ and $A$ are given by:
{\setlength\arraycolsep{2pt}
\begin{eqnarray}
\Delta&=&r^2-2Mr+a^2\label{Delta} \\
\Sigma&=&r^2+a^2\cos^2\theta\label{Sigma}  \\
A&=&(r^2+a^2)^2-a^2\Delta\sin^2\theta.\label{A}
\end{eqnarray}} 
The null geodesics are given by the tangent vector components:

{\setlength\arraycolsep{2pt}
\begin{eqnarray}
\label{geodeqt}
\dot t&=&(\Delta\Sigma)^{-1}(A\gamma-2Mar\ell) \\
\dot\theta&=&\pm \Sigma^{-1}\left[L+a^2\gamma^2\cos^2\theta-\frac{\ell^2}{\sin^2\theta}\right]^{1/2} \\
\dot\phi&=&(\Delta\Sigma)^{-1}\left[2Ma\gamma r+\frac{\ell}{\sin^2\theta}(\Sigma-2Mr)\right]\\
\dot r&=&\pm\Sigma^{-1}\left[ (2M\gamma r-a\ell)^2+\Delta(r^2+2Mr-L)\right]^{1/2}\label{geodeqr}
\end{eqnarray}}
where dot means derivative  with respect to a real parameter 
along the orbit and 
$L$, $\ell$ and $\gamma$ are constants of the motion; the parameter $L$ arises from the separability of the 
Hamilton-Jacobi equation  
in the metric (\ref{metric}) and is related in a non trivial way to the (square of) total angular momentum 
of the photon (de Felice and Preti, 1999), the parameter $\ell$ 
is the photon's azimuthal angular momentum and $\gamma$ is the photon's total energy. In what follows we shall 
introduce the new parameters $\lambda\equiv \ell/\gamma$ and $\Lambda\equiv L/\gamma^2$.

Equation (\ref{geodeqt}) shows that null geodesics exist which partially run in a time-reversal 
regime, namely with $\dot t<0$. Condition $\dot t\leq 0$ will be consistent with
the light-like character of the orbit only if 
$A<0$, namely in the $r<0-$sheet of metric (\ref{metric})
\footnote{The same argument holds true for time-like curves.}. 
From (\ref{A}) and (\ref{geodeqt}) we can state that a light signal will move in the time reversal regime if 

$i$) - it is of the vortical type (de Felice and Calvani, 1972);

$ii$) - its latitudinal angle $\theta$ satisfies the condition:
\be\label{cv}
\sin^2\theta>\frac{(r^2+a^2)^2}{a^2\Delta}-\frac{2Mr\lambda}{a\Delta}\equiv \sin^2\theta_{c.v.}
\ee
where the subscript $(c.v.)$ stands for {\it chronology violation}. 
Clearly  function 
$\sin^2\theta_{c.v.}(r;\lambda)$
identifies a region in the $(\sin^2\theta-r)-$plane whose very existence and {\it size} depend 
on the orbit itself. In particular it is 
easy to see that the smaller is $\lambda$ in the range $0<\lambda<a$ the larger is the extent 
of the $(c.v.)-$region.
Evidently, being $\lambda=\ell/\gamma$, all high energy photons with finite azimuthal angular momentum 
will have a small $\lambda$ and therefore have higher probability to go through the time reversal regime.
Condition (\ref{cv}) is  not sufficient to set up a CTM; the photon needs to encounter a turning point 
from where it can travel back to the $r>0$ universe after a sufficient recovering 
of the lost (coordinate) time.

A vortical orbit is confined between two values of the latitudinal angle $\theta$; 
in the case of photon 
orbits these angles are given by
\be
\label{theta} 
\sin^2\theta_\pm=\frac{\Lambda+a^2\pm[\Lambda+a^2)^2-4\lambda^2a^2]^{1/2}}{2a^2}
\ee
hence condition (\ref{cv}) will be satisfied only if at least one of the hyperboloids 
$\theta=\theta_\pm$ 
 as in (\ref{theta}) crosses the corresponding  $(c.v.)-$region. 
This circumstance takes place if 
\be
\label{Ltheta}
\Lambda=-a^2-\frac{\lambda^2a\Delta}{\Psi}-\frac{a\Psi}{\Delta}\equiv\Lambda_\theta
\ee
where
\be
\Psi\equiv 2M\lambda r-\frac{(r^2+a^2)^2}{a}.
\ee
Turning points are met where $\dot r=0$ and  this is assured when 
\be
\label{Lr}
\Lambda=\frac{(2Mr-a\lambda)^2}{\Delta}+r^2+2Mr\equiv\Lambda_r.
 \ee
A comparison of (\ref{Ltheta}) with (\ref{Lr}) shows that (see figure 1)
\be
\Lambda_\theta\leq \Lambda_r,\qquad (r\leq 0)
\ee
where the equality sign holds identically when $\lambda=0$ and only at $r=0$  when $\lambda^2=\Lambda$.  

\begin{figure}
\typeout{*** EPS figure 1}
\begin{center}
\includegraphics[scale=0.4]{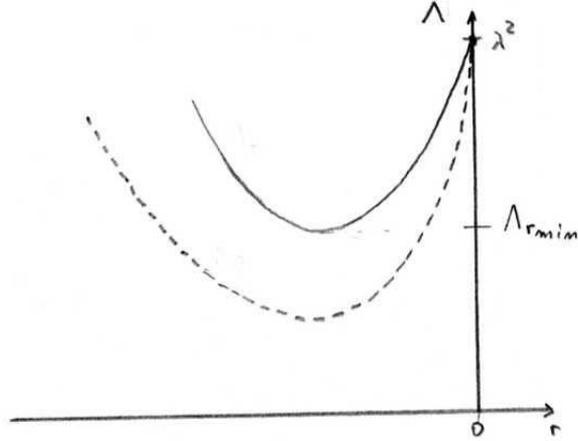}
\end{center}
\caption{Plot of the functions $\Lambda_r$ (solid line) and  $\Lambda_\theta$ (dotted line)
for a general value of $\lambda$ with $0<\lambda<a$.} 
\label{grbfig1}
\end{figure}

In the general case of $\lambda\not=0$ a CTM is 
actually set up  if condition (\ref{cv}) is satisfied together with
$\Lambda\geq\Lambda_{r_{min}}$
 where $\Lambda_{r_{min}}$ is a minimum of $\Lambda_r$ given by:
\be
\Lambda_{r_{min}} = \frac{1}{(M-r_{min})^2}[a^2(r_{min}+M)^2+2r^2_{min}(r_{min}^2-3M^2)];
\ee
here $r_{min}$ is the only negative solution of:
\be
\lambda=\frac{1}{a(M-r)}[a^2(M+r)+r^3-3Mr^2].
\ee
In particular, as pointed out by  de Felice and Calvani (1979), the existence of a minimum 
of the function $\Lambda_r$ 
assures that photons with parameters 
\be
0<\lambda<a\qquad \lambda^2>\Lambda\approx \Lambda_{r_{min}}
\ee
will move on time reversed, spatially open and almost stationary loops at an average distance $r_{min}$ 
from the singularity 
(in the $r<0$ sheet; see figure 1) before moving back to positive infinity again. 
These are the prerequisites of a Cosmic Time Machine. 

\section{The time trap}

For a light signal to move on a time reversed trajectory, the light cone must be {\it deformed}
in such a way that its future pointing generators propagate light signals into the 
local coordinate past, namely with the coordinate time decreasing. The light cone structure can 
be seen  explicitely in the Kerr naked singularity solution. From (\ref{geodeqt}) 
 and (\ref{geodeqr}) it follows that the light cone generators in the ($ct-r$)-plane satisfy the 
equation:
 \begin{equation}
 \label{generator}
 \frac{dt}{dr}=\pm\frac{a^2(\sin^2\theta_{c.v.}-\sin^2\theta)}{\left[(2Mr-a\lambda)^2+
\Delta(r^2+2Mr-\Lambda)\right]^{1/2}}
\end{equation}
where $\sin^2\theta_{c.v.}$ is given by (\ref{cv}). As $\theta\to\theta_{c.v.}$ from below, 
namely with $\theta<\theta_{c.v.}$,
  $dt/dr$ decreases untill it vanishes for {\it both} outgoing and ingoing generators.
At this moment the light cone is {\it fully} open with respect to the coordinate time axis; 
as $\theta$ increases further so that
 $\theta>\theta_{c.v.}$, the light cone shrinks again but with the local future reversed with respect 
to the coordinate time (see figure 2). 
As it will be discussed in a separate paper (de Felice and Preti, 2006) the opening of the light cone 
is a manifestation of the repulsive character 
of the space-time; indeed  this is the property of Kerr space-time nearby the ring singularity.

\begin{figure}
\typeout{*** EPS figure 2}
\begin{center}
\includegraphics[scale=0.4]{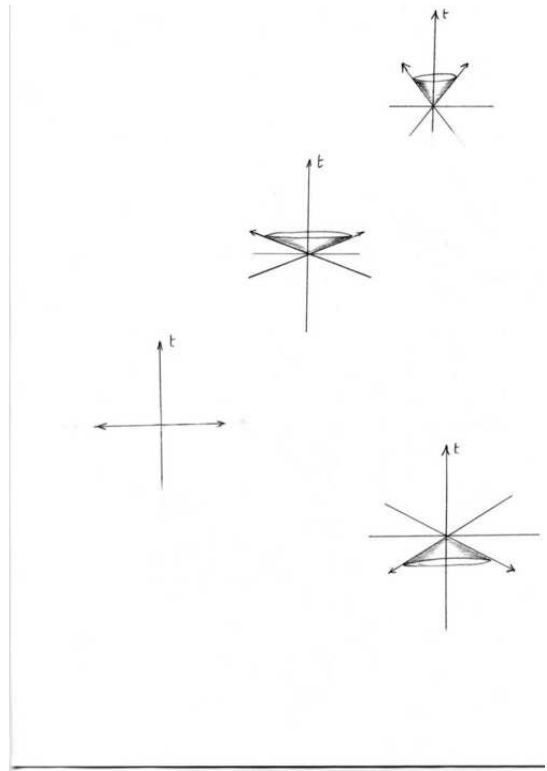}
\end{center}
\caption{ The light cone structure approaching the ring singularity in the $r<0$-sheet of the metric}
\label{grbfig2}
\end{figure}

The light trajectories which have the necessary prerequisites to become time-reversed 
act as time-traps since they force light signals to travel back in the coordinate time 
(see figure 2).
Eventually the light signals are bounced back at a tunrning point ($\dot r=0$) and that happens, as stated,
 if $\Lambda\geq \Lambda_{r_{min}}$; the coordinate time that a light signal recovers before 
going back to positive infinity depends on how close $\Lambda$ is to $\Lambda_{r_{min}}$. If 
$\Lambda\approx \Lambda_{r_{min}}$ 
 the light signal may loop on  a spatially quasi-circular orbit 
($r\sim r_{min}$) untill the coordinate time reaches reaches the value when the singularity first formed. 
At this moment the singularity is of the most general type; we expect, in fact, 
that a naked singularity becomes  of a Kerr type only when it is about to 
decay to a Kerr black hole. Although it is difficult to model the light cone 
structure nearby a  singularity at its onset, the occurrence of time inversion in its vicinity 
is assured by a general theorem as we said and will better illustrate next.

\section{A cosmic burst}
The connection between a naked singularity and a Cosmic Time Machine  
has been established in general by Clarke and de Felice (1984) with 
a theorem (theorem II of that paper). The main result of that theorem states: 
{\it if there is a naked 
singularity which satisfies Newman's strong curvature condition (Newman, 1983) and 
exists arbitrarily far into the future of a set of  initial regular data,
then violation of strong causality occurs arbitrarily close to future null infinity.
Thus a Cosmic Time Machine is naturally implied}.
This {\it pathology}
cannot be cured by any quantum correction to the classical theory of relativity 
before the singularity forms, because the very source of this peculiar behaviour is not the 
singularity itself but rather the space-time nearby it. Here, in fact, 
light cones permit non space-like trajectories to
run backwards with respect to the coordinate time causing the
local causal future to overlap with what would have been the causal past 
in a flat space-time.
This effect, which is induced by gravity, occurs (whenever it does) in a finite 
domain surrounding the singularity; this region will be termed {\it kernel} 
of a Cosmic Time Machine.
A space-time which is also a CTM must satisfy basic physical requirements. 
First the matter source which eventually evolves to a singularity must 
satisfy the energy conditions so to avoid 
quite arbitrary geometries as possible space-times. Then
the space-time solution must admit a regular flat (past and future) infinity 
for the definition of a CTM to make sense.
We shall now illustrate what are the observational implications of a CTM did it 
arise somewhere in the Cosmos. 

Let a coordinate time $t$ be chosen so to coincide 
with the proper-time of an observer at a positive infinity. 
Consider two events in a CTM-kernel being one to 
the (causal) future of the other (two subsequent flashes from the same light gun, 
say); then, being in a CTM-kernel, there exist 
light rays from these events which propagate backwards with respect
to the local time coordinate  untill they leave the kernel and escape to 
positive null infinity. If we allow for the existence of photon orbits which {\it spatially} 
loop around the singularity before leaving the CTM-kernel, it may well happen that these light 
rays leave the kernel at about the same value of the $t$ coordinate and therefore reach infinity 
at about the same value of $t$ as well.  But at flat infinity, the $t$ coordinate 
is also the proper-time of a stationary observer hence the latter would see the two events almost 
simultaneously on her (his) clock. If we extrapolate this example to all the 
events which are to the future
of any given one in a CTM-kernel, we infer that in a Cosmic Time Machine
the entire causal future development of a given domain within its kernel 
may be seen by a distant observer at the same time. 
Evidently this property makes a CTM potentially a source of an arbitrary strong burst. 

Impulsive cosmic events combine two main puzzling features, 
namely an extremely short time of emission (order of a second) and a very high energy fluence.
The main challenge therefore is to find a unique mechanism which allows at once for both properties.
The most impressive examples of the above type of events are the Gamma Ray Bursts 
(Kluzniak and Ruderman, 1998; van Putten, 2001; Piran, 2004 and references therein). 
The total energy emitted can be as 
high as $10^{54}\, ergs$, mostly concentrated in a pulse as short as a second. 
This amount of energy appears much more stunning if we think to it as being the 
energy emitted in a second-long pulse by $10^{10}$ galaxies each made of $10^{11}$ 
Sun-like stars, each emitting at a rate of $\sim 10^{33}\, ergs/sec.$, concentrated  
in a region probably smaller than a galactic core!

There are several models which provide reasonable explanations for these high 
energy events; most of them however suffer of some kind of incompleteness due to 
the rich and complicated morphology of those sources. 
Nevertheless, whatever process is considered, the common starting point is  
gravitational collapse; this may trigger 
a supernova explosion or just provide a black hole which will be the main engine for the 
burst production.

Here I envisage a completely new scenario based on the hypothesis that what we 
believe to be a black hole is on the contrary a generic strong curvature naked singularity sitting 
inside a CTM-kernel.
Since  Cosmic Time Machines involve astronomical objects, they allow one to make
predictions which could in principle be confronted {\it here-and-now} 
with observations. 
While in the kernel, in fact, the coordinate time decreases and, as we said, it may 
 reach the value when the singularity formed. 
At this time the conditions for a time trap did not yet develop  and therefore all the photons
would only propagate to the coordinate future again (coordinate time $t$ increasing) leaving the 
region nearby the singularity just formed
and leading to a  burst of radiation as seen at far distance.
As illustrated beforehand, the light cone goes from a complete inversion with respect to the 
coordinate time inside the kernel
when the local future corresponds to a decreasing $t$, to a marginal time inversion at the
kernel boundary where the future pointing light cone generators have an almost stationary $t$. 
Because of this the internal past and future are mixed 
up and can be seen at infinity over the whole time interval during which the singularity is visible.  

We can plausibly think of a  situation where an accretion disk sits around a (spinning)
naked singularity. Let a substantial part of the emitted radiation enter the kernel and 
be funneled, at least part of it, into spatially quasi-circular orbits along which light 
cones allow for local time reversed time-like or null trajectories. 
Furthermore let accretion cause an energy output of about
 $10^{40} ergs/sec$ corresponding to a moderate quasar-like object  
shining for 
some $10^{9}$ years ($\sim 10^{16}$ seconds) untill the naked singularity decays 
close to a black hole state becoming invisible to distant observers 
\footnote{ Indeed this phenomenon may repeat itself and so will do  
the effects which are here discussed.}. 
If a thiny fraction  of the emitted radiation, $\zeta = 1\%$
 say, propagates to the local future along the time-reversed orbits that we have shown to exist, 
it will likely reach the {\it bottom} of the kernel and  leave it at the same value of the 
$t$ coordinate as result of the local light cone opening.
Then an observer at infinity would  see the integrated energy of $10^{54} 
ergs$ almost at the same time.  

One can argue that the amount of radiation which was driven by the time-trap to the  bottom 
of the kernel would give rise, before flowing out, to an energy condensate
capable to alter the background geometry. Even if it is only a small fraction ($\zeta$ as said) of 
the total radiation 
emitted by the cosmic source in its life-time, 
the curvature produced by this energy concentration may be such to turn the naked singularity 
into a black hole hiding the phenomenon on the start. This may certainly be a possibility, however 
an amount of radiation of $10^{54}ergs$ as in the previous example corresponds to a gravitational source of 
$\sim 10^{33} g$, namely about one solar mass. This may be negligible if we think to a main source
singularity of $10^{8}-10^{9} M_\odot$. Moreover since we conceive a situation where most of the time-trapped
radiation is confined on spatially quasi-circular orbits
in the innermost part of the kernel, then the contribution to the geometric curvature would come 
from a rotating energy current; this would strenghten rather than weaken  
the naked singularity condition. 

Evidently the longer a naked singularity lasts as such the more luminous will be the burst because 
longer is the future development which will be "compressed" by the time inversion and therefore
more are the photons which will contribute to the prompt emission. This {\it mechanism}   
may lead to undesirable bursts of infinite intensity!  Naked singularities however appear to prevent 
this circumstance.
It is well established that
 a naked singularity  decays to a black hole. In this case, the instability of a 
Kerr black hole inner horizon 
leads to the insurgence of a "larger" singularity which will cancel any kernel structure around the 
central spinning singularity.
Even if the transitin to a black hole takes place only asymptotically, we know (Calvani and de Felice, 1978) 
that 
a Kerr naked  singularity has a "memory of the last horizon"; this manifests itself with a rather peculiar
concentration of stable spherical 
null orbits around the spatial surface $r=M$ which is the spatial location of the {\it extreme} 
Kerr black hole horizon. This concentration increases as one approaches the black hole state untill 
it creates a sort of radiation layer which will effectively prevent the reach of the CTM-kernel 
 stopping any time machine activity.

The opening of the light cone generators inside a CTM-kernel ranges from a complete reversal 
with respect to the local time axis to a marginally complete opening nearby the kernel boundary. 
It is reasonable to expect that part of the radiation emitted by the source will leave 
the kernel before it reaches its bottom and therefore at a value of the coordinate time larger 
than that when the bulk of radiation is emitted from the bottom of the kernel. 
The radiation which leaks out from the kernel  boundary will then reach the distant observer 
at a later time with respect to the main burst. This may account for the afterglows observed 
in some of the impulsive sources. 
The latters, like 
Gamma Ray Bursts for example, have a reach phenomenology and their emission properties show correlations
(Ghisellini, 2004; Ghirlanda et al., 2004a, 2004b, 2005; Piran, 2004 and references therein).  
Although the proposed scenario does not allow for definite predictions yet, we can expect obvious 
correlations.
       
 As an example consider a 
Kerr naked singularity of total mass $M$ and rotation 
parameter $a=M(1+\beta)$ where $\beta\ll 1$; because of accretion, this singularity  
will have a life time, before decaying 
to a black hole, given approximately by (de Felice, 1975):
\begin{equation}
 T\approx 1.5\times 10^4\frac{M_\odot}{M}\frac 1{\rho(1+\beta)}\,years
 \end{equation}
Here $\rho$ is the density of matter which accretes on the singularity. 
The life time then critically depends on the  factor $(M\rho)^{-1}$.
For sake of illustration, a $10^9M_\odot$ naked singularity will last for $10^8$ years, 
as in the previous example, if it is surrounded by accreting material of density
$\rho\approx 10^{-13}\, g\,cm^{-3}$, a value which appears compatible with 
what one could have in active galactic nuclei. 
In this scenario then we expect that the total 
luminosity of the impulsive emission goes as $L\sim  \zeta(M\rho)^{-1}$.
Moreover, the longer a naked singularity is visible to distant observers the longer one expects the 
afterglows to last. Hence a correlation such as more luminous burst being followed by longer afterglows 
is expected.

Evidently the survival of the above conjecture about the nature of impulsive sources depends
on the possibility to be falsified by more definite observational constraints; this however is a 
challenge for the future. 

\section{Conclusions}
If naked singularities exist in the Cosmos as predicted by general relativity
then they may give rise to a Cosmic Time Machine. In this case, in fact, the 
singularity could likely be surrounded by a space time region where the local causal future is 
time-inverted with respect to infinity. Naked singularities however 
will likely evolve close to a  black-hole state in a finite interval of coordinate time 
hence if all that happens, then we could directly observe 
 astrophysical phenomena which are observationally constrained by the 
peculiarities of a 
time machine. This may be the case of the most energetic Gamma Ray Bursts whose 
impulsive emission may just be the time integration over a finite interval of the 
local proper-time  
of  emission processes taking place in some CTM-kernel.The latter then will be sources of 
the most powerful bursts in the Universe.

\end{document}